\documentclass[12pt]{article}
\usepackage{amsfonts,amsmath,amssymb}
\usepackage{mathrsfs}
\usepackage{theorem}
\usepackage{graphics,epsfig}
\newcommand{\be}{\begin{equation}}
\newcommand{\ee}{\end{equation}}
\newcommand{\bea}{\begin{eqnarray}}
\newcommand{\eea}{\end{eqnarray}}
\newcommand{\half}{\frac{1}{2}}
\newcommand{\halfi}{\frac{i}{2}}
\newcommand{\la}{\langle}
\newcommand{\ra}{\rangle}
\newcommand{\nn}{\nonumber}
\newcommand{\eps}{\epsilon}
\newcommand{\liealg}{\mathfrak}
\theorembodyfont{\rmfamily}

          \def\cH{{\cal H}}

          \def\cW{{\cal W}}

\newcommand{\RR}{{\mathbb R}}
\def\lddots{\mathinner{\mkern1mu\raise1pt\hbox{.}\mkern2mu
\raise4pt\hbox{.}\mkern2mu\raise7pt\vbox{\kern7pt\hbox{.}}\mkern1mu}}

\def\qmbox#1{\qquad\mbox{#1}\quad}

\begin{document}

\pagestyle{empty} \setcounter{page}{0}


\strut\hfill{}

\vspace{0.5in}

\begin{center}

{\Large \textsf{Bethe Equations for a $\liealg g_2$ Model\\[5mm]
}}

\vspace{10mm}

{\large N.Cramp\'e\footnote{nc501@york.ac.uk} and
C.A.S.Young\footnote{charlesyoung@cantab.net}}

\vspace{10mm}

Department of Mathematics\\
 University of
York\\
 Heslington York\\
  YO10 5DD, United Kingdom\\

\end{center}

\vfill \vfill

\begin{abstract}
We prove, using the coordinate Bethe ansatz, the exact solvability
of a model of three particles whose point-like interactions are
determined by the root system of $\liealg g_2$. The statistics of
the wavefunction are left unspecified.  Using the properties of the
Weyl group, we are also able to find Bethe equations. It is notable
that the method relies on a certain generalized version of the
well-known Yang-Baxter equation. A particular class of non-trivial
solutions to this equation emerges naturally.

\end{abstract}

\vfill MSC numbers: 82B23, 81R12, 70H06

PACS numbers: 02.30.Ik, 03.65.Fd \vfill

\baselineskip=16pt

\newpage
\pagestyle{plain}

Using the coordinate Bethe ansatz \cite{bethe}, C.N.Yang \cite{yang} solved a
model of $n$ particles, with unspecified statistics, interacting via
contact interactions. This procedure led him to discover the celebrated Yang-Baxter equation, found also, in
the context of statistical physics, by R.J.Baxter \cite{baxter}. The
potential used in the approach of C.N.Yang is intimately linked to
the simple root system of $\liealg{sl}_n$. The generalization to the other root
systems has also been intensively studied (see for example \cite{gaudin,olpe}).
In the case of the $\liealg{so}_n$ and  $\liealg{sp}_n$ root systems, a new type of equation,
the so-called \emph{reflection equation} \cite{che,skly}, is obtained. It
plays a fundamental role in the study of integrable system with boundaries.

In this letter, we demonstrate, using the procedure of \cite{yang},
the exact solvability of a model based on the $\liealg{g}_2$ root
system. This gives rise to a generalized version of the Yang-Baxter
equation \cite{gaudin, Sutherland, che} which is peculiar to
$\liealg g_2$ amongst the simple Lie algebras and is distinct from
the usual Yang-Baxter and reflection equations. We then find the
Bethe equations for the model, which involves some interesting
subtleties.

\subsection*{The Model}
Let us consider a system of three particles with positions $x_1$, $x_2$ and $x_3$, whose interactions are specified by the Hamiltonian
\begin{equation}
\label{ham} H=-\sum_{i=1}^3 \frac{\partial^2}{\partial x_i^2}+2 g_S
\mathop{\sum_{i,j=1}}_{i\neq j}^3 \delta(x_i-x_j)+2 g_L
\mathop{\sum_{i,j,k=1}}_{i\neq j\neq k\neq i}^3 \delta(x_k-\half(x_i+x_j)).
\end{equation}
Here $g_S$, $g_L$ are real parameters characterizing the strength of two types of interactions. Physically, the $g_S$ term is the
usual contact term -- specifying how particles interact when they collide -- while the $g_L$ term may be thought of as describing
a contact interaction between each particle and the centre of mass of the remaining pair.
We wish to solve the spectral problem
\begin{equation}
\label{blem} H \phi(x_1,x_2,x_3) = E \phi(x_1,x_2,x_3).
\end{equation}

The motivation for the Hamiltonian (\ref{ham}) is that it is related
to the root system of the exceptional Lie algebra $\liealg{g}_2$
\cite{olpe}, just as the model of $n$ particles with purely contact
interactions is related to that of $\liealg sl_n$: the root system
of $\liealg g_2$ is
\begin{equation}
\label{root}
\Delta=\{\epsilon_i-\epsilon_j,\epsilon_i+\epsilon_j-2\epsilon_k|1\leqslant
i\neq j\neq k\neq i\leqslant 3\},
\end{equation}
where $\{\epsilon_i\}$ an orthonormal basis of $\RR^3$. In terms of
these roots, the Hamiltonian reads as
\begin{equation}
\label{ham1} H=-\sum_{i=1}^3 \frac{\partial^2}{\partial x_i^2}+
\sum_{\alpha\in \Delta} g_{\la\alpha,\alpha\ra}~ \delta(\la \alpha,x\ra)
\end{equation}
where $x=\sum_ix_i\eps_i$, $g_2=g_S$, $g_6=g_L$  and $\la\cdot,\cdot\ra$ is the usual scalar product.

\subsection*{Coordinate Bethe Ansatz}
To translate the problem into one which may be solved by the coordinate Bethe ansatz, it is useful to consider the
Weyl group associated to the Lie algebra $\liealg g_2$ \cite{gaudin}. To
each root $\alpha\in \Delta$ is associated a reflection in the
hyperplane $\cH_{\alpha}$ perpendicular to $\alpha$:
\begin{equation}
s_{\alpha}(\beta)=\beta -2~\frac{\la\alpha,\beta\ra}
{\la\alpha,\alpha\ra}~\alpha\;.
\end{equation}
The set of these reflections, together with the identity $I$, form
the Weyl group of $\liealg g_2$, which maps the set of roots to
itself. The group is generated by the reflections in the two simple
roots \be \alpha_1=\epsilon_1-\epsilon_2
\qmbox{and}\alpha_2=\epsilon_2+\epsilon_3-2\epsilon_1,\ee which we
denote by, respectively, $T$ and $R$. The relations obeyed by these
generators
\begin{eqnarray}
\label{defD6} \left(T\right)^2=I \qmbox{,} \left(R\right)^2=I
\qmbox{and} \left(TR\right)^6=I\;,
\end{eqnarray}
completely specify the group, which is isomorphic to the dihedral group $D_6$.

\begin{figure}[htb]
\begin{center}
\epsfig{file=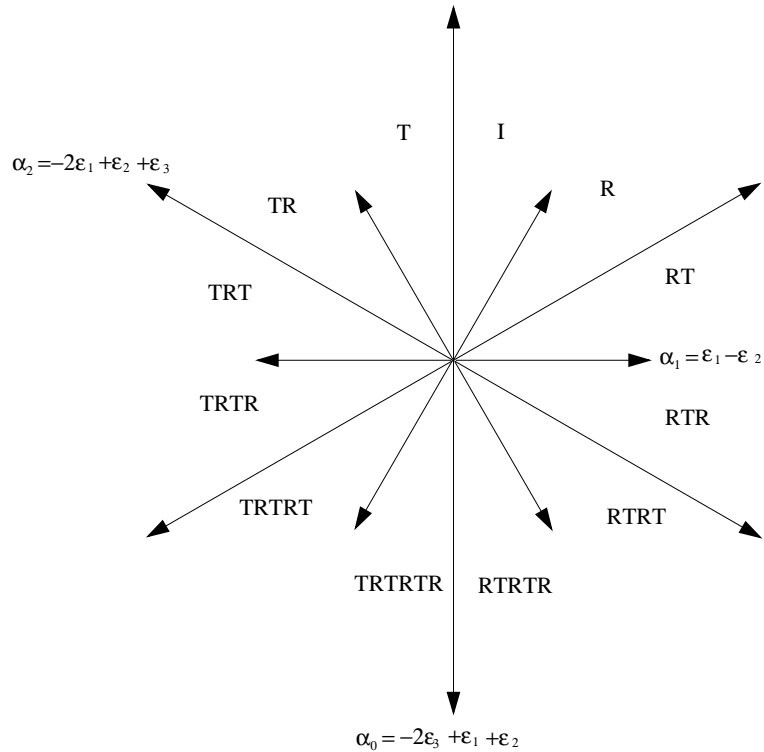,width=10cm}
\end{center}
\caption{Root system of $\liealg{g}_2$, showing the labelling of Weyl chambers.
\label{Fig:roots}}
\end{figure}

The twelve hyperplanes $\cH_{\alpha}$ define twelve domains, called Weyl chambers, in $\RR^3$. These
chambers are characterized by
\begin{equation}
\label{region} \cW_{s_{\alpha}}:~ 0<
x_{s_{\alpha}\!1}-x_{s_{\alpha}\!2}
<x_{s_{\alpha}\!3}-x_{s_{\alpha}\!1}\,,
\end{equation}
where $x_{s_{\alpha}\!i}$ is the $i^{th}$ component of the vector
$\big(s_{\alpha}\big)^{-1}(x)$. Let us remark that conditions
(\ref{region}) are equivalent to
\begin{equation}
\cW_{s_{\alpha}}:~x_{s_{\alpha}\!2}<x_{s_{\alpha}\!1}\qmbox{and}
x_{s_{\alpha}\!1}< \frac{x_{s_{\alpha}\!2}+x_{s_{\alpha}\!3}}{2}.
\end{equation}
In each Weyl chamber, the three particles are in a given order, as
are the centre of mass of the two extreme particles and the middle
particle. The importance of the definition of the Weyl group in this
context is two-fold \cite{gaudin}: the hyperplanes, $\cH_{\alpha}$,
are the domains of the configuration space where the interactions
take place; and the action successively of $R$ and $T$ allows us to
describe, starting from one region $\cW_{s_\alpha}$, the eleven
other regions. (See figure \ref{Fig:roots}.) Using these two
properties, the spectral problem (\ref{blem}) may be written
equivalently as a free Hamiltonian, for $(x_1,x_2,x_3)\notin
\{\cH_{\alpha}\;|\;\alpha\in \Delta\}$,
\begin{equation}
\label{free} -\sum_{i=1}^3 \frac{\partial^2}{\partial
x_i^2}\phi(x_1,x_2,x_3) =E\phi(x_1,x_2,x_3)
\end{equation}
and boundary conditions on hyperplanes $\cH_{\alpha}$ ($\alpha\in
\Delta$)
\begin{eqnarray}
\label{bound1} \phi|_{ \la \alpha,x \ra =0^+}= \phi|_{\la \alpha,x\ra=0^-}
~~\mbox{and}~~
\la\alpha,\nabla\ra \phi|_{\la \alpha,x\ra =0^+}=\left(\la \alpha,\nabla\ra
+2g_{\la \alpha,\alpha\ra }\right)\phi|_{\la\alpha,x\ra=0^-}
\end{eqnarray}
where $\nabla=\sum_i \eps_i \frac{\partial}{\partial x_i}$. The
equivalence between the $\delta$-potentials in the Hamiltonian
(\ref{ham1}) and these boundary conditions (\ref{bound1}) is
well-known (see for example \cite{gaudin}).

We now determine the eigenfunctions of the Hamiltonian by solving
equation (\ref{free}) with boundary conditions (\ref{bound1}). We
make the following ansatz for $\phi$: in the region $\cW_Q$ (with
$Q\in D_6$), the eigenfunction of the Hamiltonian is written as
follows \be \phi_Q(x_1,x_2,x_3)=\sum_{P\in D_6} \exp(i \la k_P, x_Q
\ra )A_P(Q).\ee This ansatz is similar to the one suggested by
H.Bethe \cite{bethe}: the only difference is that here the sum is
over a dihedral, rather than permutation, group. These
eigenfunctions obviously satisfy relation (\ref{free}) with
\begin{equation}
E=\sum_{i=1}^3 k_i^2\;.
\end{equation}
We need to determine the parameters $A_P(Q)$ present in the ansatz
so that boundary conditions (\ref{bound1}) are satisfied. Because of
definition (\ref{region}) of the region $\cW_Q$ ($Q\in D_6$), the
eigenfunction $\phi_Q$ adjoins two boundaries:
$x_{Q1}=\frac{x_{Q2}+x_{Q3}}{2}$ and $x_{Q1}=x_{Q2}$. The boundary
conditions imply the following constraints between the vectors $A_P$ (whose components are
$A_P(Q)$, $P,Q\in D_6$):
\begin{eqnarray}
\label{AA} A_{PR}=B(\la k_P, \alpha_2 \ra )A_P\qmbox{and}
A_{PT}=Y( \la k_P, \alpha_1 \ra )A_P
\end{eqnarray}
where
\begin{eqnarray}
\label{defBY} B(k)=\frac{k\widehat{R}+ig_L}{k-ig_L}\qmbox{and}
Y(k)=\frac{k\widehat{T}+ig_S}{k-ig_S}\;
\end{eqnarray}
and the operators $\widehat{R}$ and $\widehat{T}$, which provide a
realization of the group $D_6$, are defined by
\begin{eqnarray}
\widehat{R}A_P(Q)=A_P(QR)\qmbox{and}\widehat{T}A_P(Q)=A_P(QT)\;.
\end{eqnarray}

One can now use relations (\ref{AA}) to calculate recursively all
the $A_P$ starting from, for example, $A_I$.
However, due to the relations (\ref{defD6}) satisfied by the
generators of $D_6$, the following consistency relations appear
between the operators (\ref{defBY})
\begin{eqnarray}
\label{unitarity}
&&B(k)=\left(B(-k)\right)^{-1}\qmbox{,}Y(k)=\left(Y(-k)\right)^{-1}
\end{eqnarray}
and
\begin{eqnarray}
\label{gYBE} &&\hspace{-2cm}Y(k_1-k_2)B(2k_1-k_2-k_3)Y(k_1-k_3)
B(k_1+k_2-2k_3)Y(k_2-k_3)B(2k_2-k_1-k_3)\nn\\
&&\hspace{-1cm}= B(2k_2-k_1-k_3)Y(k_2-k_3)B(k_1+k_2-2k_3)
Y(k_1-k_3)B(2k_1-k_2-k_3)Y(k_1-k_2)
\end{eqnarray}
Using solely that $\widehat{R}$ and $\widehat{T}$ satisfy relations
(\ref{defD6}), one can verify by direct computation that these
equations hold, finishing our argument about the exact solvability
of the $\liealg g_2$ model. The relations (\ref{unitarity}) are the
usual unitarity relations whereas (\ref{gYBE}) is a generalization
of the Yang-Baxter equation \cite{gaudin,Sutherland,che}. We discuss
the implications of its appearance briefly below.

\subsection*{Bethe Equations}
In order to find the Bethe equations we suppose that the three
particles live on a circle of finite circumference $2L$. One must
specify carefully how the particles interact in this case. As on the
infinite line, there are interactions due to direct collisions of
particles. But, since on a circle there is no preferred notion of
which particle lies on the left, on the right, or in the middle, it
is now most natural to assume that \emph{each} of the three
particles interacts with the midpoint of the remaining pair. What is
more, for each pair there are really two ``midpoints'', and we
assume that the third particle interacts with both. (The midpoint of
two particles would otherwise jump discontinuously as they pass
opposite points, which seems physically unappealing.)

\begin{figure}[htb]
\begin{center}
\epsfig{file=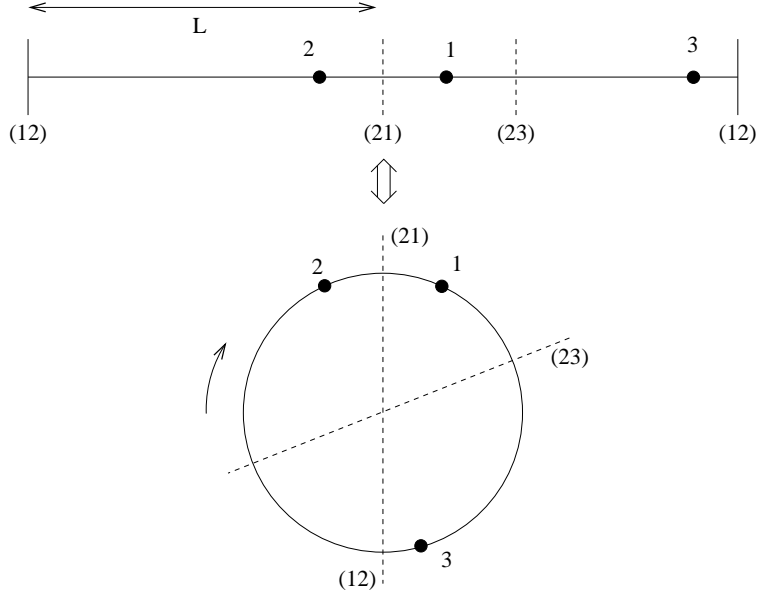,width=10cm}
\end{center}
\caption{A configuration in region $\cW_I$. (The midpoint line
$(13)$ is not shown.)\label{Fig:circline}}
\end{figure}

To make contact with the preceding section we need a prescription to
identify the set of possible configurations of the particles on a
circle with the group $D_6$. We make the following choice. We
``cut'' the circle at the point opposite the mid-point of the
closest pair of particles (call them $A$ and $B$ -- of course, the
particles cannot be equally spaced for then each would lie on the
mid-point of the other two). We then unwrap the circle to get an
interval of length $2L$, and the positions of the particles on this
interval correspond to a unique element of $Q\in D_6$, just as in
the previous section. (Figure \ref{Fig:circline} illustrates an
example in which $Q=I$.)

By construction, neither $A$ nor $B$ can lie at an endpoint of this
interval (in fact, neither can be closer than $2L/3$ to an endpoint)
so we need only specify the boundary conditions for the remaining
particle, $C$.\footnote{Note also that the prescription ensures that
-- to take the example in figure \ref{Fig:circline} -- $1$ reaches
the ``internal'' midpoint $(23)$ \emph{before} $2$ has a chance to
reach the ``external'' midpoint $(13)$. So the boundary condition on
particle $C$ ($3$, in this case) is genuinely the only new
condition, not present in the model on the line.} On examining the
definitions of the various regions, one sees that $C$ can reach only
the right endpoint for all configurations corresponding to \be Q \in
\mathcal B^+ := \left\{ I,\,\, T,\,\, RTR,\,\, TRTR,\,\,
RTRT,\,\,TRTRT\right\}  \ee and that it can reach only the left
endpoint for all configurations corresponding to \be Q \in \mathcal
B^+ W = \left\{ RTRTR,\,\, (RT)^3,\,\, TR,\,\, R, \,\,TRT,\,\, RT
\right\} \ee where we define $W=RTRTR$. More specifically, when one
starts in a region $\cW_Q$, with $Q\in \mathcal B^+$, and moves
particle $C$ through the right boundary, one reaches the region
$\cW_{QW}$. As it crosses the boundary, $C$ interacts with the
``opposite'' (on the circle) midpoint of $A$ and $B$, so the
conditions are continuity with a jump in the first derivative, just
in (\ref{bound1}) above. Explicitly, one finds \be \phi|_{\la
\alpha_0,x_Q\ra= -2L^-} = \phi|_{\la \alpha_0, x_Q \ra = + 2L^- }\ee
and \be \la \alpha_0 , \nabla_Q \ra \phi|_{\la \alpha_0,x_Q\ra=
-2L^-} = \left( \la \alpha_0, \nabla_Q \ra + 2g_L\right)
                       \phi|_{\la \alpha_0, x_Q \ra = + 2L^-  },\ee
where $\alpha_0=\epsilon_1+\epsilon_2-2\epsilon_3$ is the lowest
root. Note that $W=RTRTR$ is the reflection in the plane orthogonal
to $\alpha_0$, so that \be \la \alpha_0, x_Q \ra = - \la \alpha_0,
x_{QW} \ra \ee and hence one can replace $Q$ with $QW$ in these
boundary conditions without altering their content -- as expected,
since this is supposed to describe the boundary between $\cW_Q$ and
$\cW_{QW}$. In fact the boundary conditions may be re-written most
symmetrically as \be \phi|_{\la \alpha_0,x_Q\ra= -2L^-} = \phi|_{\la
\alpha_0, x_{QW} \ra = - 2L^-  }\ee and \be \la \alpha_0 , \nabla_Q
\ra \phi|_{\la \alpha_0,x_Q\ra= -2L^-} = \left( \la \alpha_0,
\nabla_Q \ra + 2g_L\right)
                       \phi|_{\la \alpha_0, x_{QW} \ra = - 2L^-  }.\ee

Substituting the Bethe ansatz for $\phi$ (in region $\cW_Q$ on the
left and region $\cW_{QW}$ on the right), one has, from the first
boundary condition,
 \be \sum_{P\in D_6} A_P(Q)  e^{i \la k_P, x_Q\ra
}|_{\la \alpha_0,x_Q\ra= -2L^-}
   =\sum_{P\in D_6} A_P(QW) e^{i \la k_P, x_{QW}\ra}|_{\la \alpha_0,x_{QW}\ra= -2L^-}.\ee
Now the vectors $n=\frac{1}{\sqrt{3}}(1,1,1)$, $\alpha_1$ and
$\alpha_0$ are orthogonal, and hence
\be \la k_P,x_Q \ra = \la k_P,
n \ra \la n, x_Q \ra + \half \la k_P, \alpha_1 \ra \la \alpha_1 ,
x_Q \ra
 + \frac{1}{6} \la k_P, \alpha_0 \ra \la \alpha_0 , x_Q \ra;\ee
furthermore $W$ fixes $n$ and $\alpha_1$ and inverts $\alpha_0$.
Thus
\bea && \sum_{P\in \mathcal B^+} e^{i \la k_P, n \ra \la n, x_Q
\ra + \halfi \la k_P, \alpha_1 \ra \la \alpha_1 , x_Q \ra}
            \left( A_P(Q) e^{ - \frac{iL}{3} \la k_P, \alpha_0 \ra} + A_{PW}(Q) e^{ + \frac{iL}{3} \la k_P, \alpha_0 \ra}\right)\nn\\
&=& \sum_{P\in \mathcal B^+} e^{i \la k_P, n \ra \la n, x_Q \ra + \halfi \la k_P, \alpha_1 \ra \la \alpha_1 , x_Q \ra}
            \left( A_P(QW) e^{ - \frac{iL}{3} \la k_P, \alpha_0 \ra} + A_{PW}(QW) e^{ + \frac{iL}{3} \la k_P, \alpha_0 \ra}\right),\eea
which can hold for general $x$ only if, for all $P\in \mathcal B^+$,
\be A_P(Q) e^{ - \frac{iL}{3} \la k_P, \alpha_0 \ra} + A_{PW}(Q) e^{ + \frac{iL}{3} \la k_P, \alpha_0 \ra}
   = A_P(QW) e^{ - \frac{iL}{3} \la k_P, \alpha_0 \ra} + A_{PW}(QW) e^{ + \frac{iL}{3} \la k_P, \alpha_0 \ra}.\ee

Meanwhile, from the boundary condition on the first derivative of
$\phi$ one finds, by very similar reasoning, that
\bea && i \la k_P
, \alpha_0 \ra A_P(Q) e^{ - \frac{iL}{3} \la k_P, \alpha_0 \ra}
 -  i\la k_{P} , \alpha_0 \ra  A_{PW}(Q) e^{ + \frac{iL}{3} \la k_P, \alpha_0 \ra}\nn\\
   &=&  \left(-i\la k_{P} , \alpha_0 \ra +2g_L\right) A_P(QW) e^{ - \frac{iL}{3} \la k_P, \alpha_0 \ra}
 + \left(i\la k_{P} , \alpha_0 \ra +2g_L\right) A_{PW}(QW) e^{ + \frac{iL}{3} \la k_P, \alpha_0 \ra}\eea
(here we used $\la k_P,x_{QW}\ra=\la k_{PW},x_Q\ra$ for the first
term on the right).

After eliminating $A_{PW}(QW)$ in the second equation using the
first, one arrives at
\be e^{\frac{2iL}{3} \la k_P,\alpha_0\ra}
A_{PW} 
     = \frac{\la k_P,\alpha_0\ra \hat W + ig_L}{\la k_P,\alpha_0 \ra - ig_L} A_P\label{apww}\ee
where $\hat W A_P(Q) = A_P(QW)$. This is true for all $P\in \mathcal
B^+$,\footnote{In fact it is, as an immediate consequence, true for
all $P\in D_6$: on setting $P=P'W$ one finds the same equation for
$P'$.} but there is some redundancy, for if (\ref{apww}) holds for
$P$ it also holds for $PT$:
\bea e^{\frac{2iL}{3} \la k_{PT},\alpha_0 \ra} A_{PTW}  &=&   e^{\frac{2iL}{3} \la k_{PT},\alpha_0 \ra} A_{PWT}  \nn\\
                                &=&  e^{\frac{2iL}{3} \la k_{P},\alpha_0 \ra} Y(\la k_{PW} ,\alpha_1 \ra) A_{PW} \nn\\
        &=&  Y(\la k_{PW} ,\alpha_1 \ra)  \frac{\la k_{P},\alpha_0\ra \hat W + ig_L}{\la k_{P},\alpha_0 \ra - ig_L} A_{P}\nn\\
    &=&  \frac{\la k_{P},\alpha_0\ra \hat W + ig_L}{\la k_{P},\alpha_0 \ra - ig_L}  Y(\la k_{PW} ,\alpha_1 \ra) A_P \nn\\
     &=& \frac{\la k_{PT},\alpha_0\ra \hat W + ig_L}{\la k_{PT},\alpha_0 \ra - ig_L} A_{PT}\eea
(the essential point is that $\alpha_0$ and $\alpha_1$ are
orthogonal, so $W=s_{\alpha_0}$ fixes $\alpha_1$, $T=s_{\alpha_1}$
fixes $\alpha_0$, and $[W,T]=0$). It therefore suffices to consider
\be
P\in \left\{ I,\quad RTR,\quad TRTR\right\}.
\ee
Using now the
relations (\ref{AA}), which come from the ``interior'' boundary
conditions, (\ref{apww}) yields three equations for $A_I(Q)$. For
example, in the case $P=I$, \be A_{W} = A_{RTRTR} = B(\la
k_{RTRT},\alpha_2\ra) Y(\la k_{RTR},\alpha_1\ra)
                          B(\la k_{RT},\alpha_2\ra) Y(\la k_{R},\alpha_1\ra)
                           B(\la k,\alpha_2\ra) A_I \label{apw}.\ee
To make the content of the resulting equations clearer, it is
helpful to define some new operators. First, for every root
$\alpha$, let $\hat s_\alpha A_P(Q) = A_P(Q s_\alpha)$ and define
\be Z_\alpha(k) = \frac{ \la k, \alpha \ra  + i g_{\la \alpha,
\alpha \ra}\hat s_\alpha }
                           { \la k,\alpha \ra - i g_{\la \alpha, \alpha\ra}},\ee
which has the property that $Z_{\alpha}(k)^{-1} = Z_{\alpha}(-k)$.
This notation is compact but rather opaque, so it is useful to
define also \be S_{12} = Z_{\epsilon_1-\epsilon_2},\quad S_{23} =
Z_{\epsilon_2-\epsilon_3},\quad S_{31} =
Z_{\epsilon_3-\epsilon_1},\ee (whose inverses we write as $S_{21}$,
$S_{32}$, and $S_{13}$) and similarly for the long roots \be
K_{23}^1 = Z_{\epsilon_2+\epsilon_3-2\epsilon_1},\quad K_{31}^2 =
Z_{\epsilon_3+\epsilon_1-2\epsilon_2},\quad
      K_{12}^3 = Z_{\epsilon_1+\epsilon_2-2\epsilon_3}.\ee

In terms of these operators we find, for $P=RTR$, $P=TRTR$ and $P=I$ respectively,
\be e^{\frac{2iL}{3} \la k,\eps_2+\eps_3-2\eps_1 \ra} A_I = \mathscr{R}_1 A_I :=
S_{21}(k) K_{13}^2(k)^{-1} K_{23}^1(k) K_{12}^3(k)^{-1} S_{31}(k) K_{23}^1(k) A_I ,\label{bethe1}\ee
\be e^{\frac{2iL}{3} \la k,\eps_3+\eps_1-2\eps_2 \ra} A_I = \mathscr{R}_2 A_I :=
K_{23}^1(k)^{-1} K_{13}^2(k) K_{12}^3(k)^{-1} S_{32}(k) K_{13}^2(k) S_{12}(k) A_I ,\label{bethe2}\ee
\be e^{\frac{2iL}{3} \la k,\eps_1+\eps_2-2\eps_3 \ra} A_I =  \mathscr{R}_3 A_I :=
K_{23}^1(k)^{-1} S_{13}(k) K_{12}^3(k) S_{23}(k) K_{13}^2(k)^{-1} K_{12}^3(k) A_I .\label{bethe3}\ee

These are the Bethe equations for the problem, and the task is to
show that the $\mathscr{R}_i$ commute. But, before this, it is
important to observe that there is further redundancy. The
exponentials on the left hand sides contain the three long roots of
$\liealg g_2$, which are of course co-planar and indeed sum to zero.
Thus there are really only \emph{two} independent equations, and,
multiplying the equations above together, we find that we must have
\be \mathscr R_1 \mathscr R_2 \mathscr R_3 A_I = A_I.\ee

To see that this is in fact true -- which is a good consistency
check -- and to verify the commutation relations $[\mathscr R_i,
\mathscr R_j]=0$, one use the following properties of $S$ and $K$:
\be S_{12} K_{12}^3 = K_{12}^3 S_{12}\ee \be K_{12}^3
\left(K_{13}^2\right)^{-1} K_{23}^1 =  K_{23}^1
\left(K_{13}^2\right)^{-1} K_{12}^3\ee \be S_{12} K_{23}^1 S_{13}
\left(K_{12}^3\right)^{-1} S_{23} K_{13}^2 =
 K_{13}^2 S_{23} \left(K_{12}^3\right)^{-1} S_{13} K_{23}^1 S_{12}.
 \label{gYBE2}\ee
These may be verified directly. (See also
\cite{Cherednik:1994ez,ChariPres,Drinfeld}.)

The reason there are only two equations -- even though there are
three momenta $k_i$ -- is that we chose to apply periodic boundary
conditions in a fashion which made no reference to any fixed point
on the circle. (It is more usual \cite{Lieb,yang,MA} to take, for
example $\phi|_{x_1=0}=\phi|_{x_1=2L}$, but in the present case the
nature of the interactions make this technically inconvenient.) Thus
the symmetry of the problem under rotations was kept manifest
throughout and the corresponding conserved quantity, the total
(angular) momentum \be P= k_1 + k_2 + k_3 = \sqrt{3} \la k , n
\ra,\ee dropped out of the calculation. But of course, as in any
quantum-mechanical system of particles on a circle, $P$ is the
generator of rigid rotations, and the invariance of the problem
under a rotation through one complete turn produces the quantization
condition \be 1 = e^{2iLP} = e^{2iL \la k,\eps_1+\eps_2+\eps_3\ra
}.\ee This, together with any two of (\ref{bethe1}-\ref{bethe3}),
gives the complete set of quantization conditions on the momenta
$k_i$.

It is interesting to note that, at least in the centre of momentum frame $k_1+k_2+k_3=0$, the equations (\ref{bethe1}-\ref{bethe3})
have the intuitive interpretation one would expect: for example, (\ref{bethe2}) becomes
\be e^{2iL k_2} A_I = K_{23}^1(k)^{-1} K_{13}^2(k) K_{12}^3(k)^{-1} S_{32}(k) K_{13}^2(k) S_{12}(k) A_I \ee
and describes the process of moving particle 2 clockwise through one complete revolution while the other particles remain fixed.
Thus, the first event is the scattering of 1 and 2 (hence $S_{12}$) followed by particle 2 interacting with the midpoint of 1 and 3
(giving $K_{13}^2$), scattering with 3 ($S_{32}$), and so on.

\subsection*{Conclusion}

\begin{figure}[htp]
\begin{center}
\epsfig{file=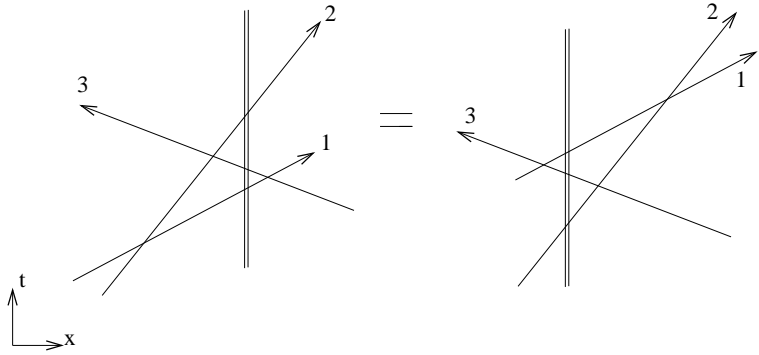,width=10cm}
\end{center}
\caption{Pictorial representation of the generalized Yang-Baxter
equation (\ref{gYBE2}) \label{Fig:gYBE}}
\end{figure}

To conclude, let us comment briefly on the generalized Yang-Baxter
equation (\ref{gYBE2}) we obtained. Like the Yang-Baxter and
reflection equations, (\ref{gYBE2}) may be represented
diagrammatically. This is shown in Figure \ref{Fig:gYBE}, where, to
simplify the picture, we restrict ourselves to the case where
$x_1+x_2+x_3=0$.
The three arrows represent the ``trajectories'' of the three particles
and the double-line that of their centre of mass. The intersection
between the two arrows $a$ and $b$ corresponds to the scattering of $S_{ab}$
of particles $a$ and $b$. The intersection between the arrow
$a$ with the double-line corresponds to the scattering $K_{bc}^a$ ($b,c\neq a$)
between $a$ and the centre of mass. Obviously,
we recover the usual representation of the Yang-Baxter equation by
removing the double-line. This occurs in the limit $g_L=0$, for then
$K_{bc}^a=1$.

Since the Yang-Baxter and reflection equations play a fundamental
role in the development of integrable models and quantum groups, it
is natural to speculate that the generalized Yang-Baxter equation
(\ref{gYBE2}) might also have interesting applications. In
particular, we hope that they will allow one to study integrable
models where the interactions between three particles are not
factorisable.


\vspace{2cm} \textbf{Acknowledgements:} NC is grateful for the
financial support of the TMR Network "EUCLID. Integrable models and
applications: from strings to condensed matter", contract number
HPRN-CT-2002-00325. CASY gratefully acknowledges the financial support
of PPARC.

\end{document}